\documentclass[epjST]{svjour}
\usepackage{graphics}
\usepackage{amsfonts}
\usepackage{psfrag}

\def\be{\begin{equation}}
\def\ee{\end{equation}}
\def\ba{\begin{eqnarray}}
\def\ea{\end{eqnarray}}

\def\R{{\mathbb{R}}}

\begin{document}

\title{Infinite disorder and correlation fixed point
in the Potts model with correlated disorder}
\author{C. Chatelain\inst{1}}
\institute{Groupe de Physique Statistique,
D\'epartement P2M,
Institut Jean Lamour (CNRS UMR 7198),
Universit\'e de Lorraine, France,
\email{christophe.chatelain@univ-lorraine.fr}}

\abstract{
Recent Monte Carlo simulations of the $q$-state Potts model with a
disorder displaying slowly-decaying correlations reported a violation
of hyperscaling relation caused by large disorder fluctuations and the
existence of a Griffiths phase, as in random systems governed by an
infinite-disorder fixed point. New simulations, directly made in the
limit of an infinite disorder strength, are presented. The magnetic
scaling dimension is shown to correspond to the correlated percolation
fixed point. The latter is shown to be unstable at finite disorder
strength but with a large cross-over length which is not accessible
to Monte Carlo simulations.}

\maketitle

\section{Introduction}
\label{intro}
Disorder is present in all experimental systems but it does not lead
to the same consequences. The study of its influence on phase transitions
is of particular interest because even a weak disorder can induce drastic
changes. First-order phase transitions are softened and may even become
continuous if the disorder is sufficiently strong or the space dimension
equal to two~\cite{ImryWortis,Aizenman}.
When the transition of the pure system is already continuous, its critical
behavior is changed by the introduction of disorder when the random
fluctuations grow faster with the system size than energy
fluctuations~\cite{Harris}. The critical behavior of the random system is
then governed by a new Renormalization Group (RG) fixed point. The fixed
point of the pure model, still present, is unstable but may cause a
cross-over at weak disorder. A clear example of a change of universality
class upon the introduction of disorder is provided by the two-dimensional
3 or 4 state Potts model, a generalization of the celebrated Ising model.
When exchange couplings are made random, new critical exponents were
measured numerically~\cite{Cardy,Cardy2,Cardy3} and shown to be in
agreement with RG calculations~\cite{RBPM-RG,RBPM-RG2,RBPM-RG3,RBPM-RG4}.
Experimentally, the phase transition of
the 4-state Potts model can be realized by the order-disorder transition
of atoms binded on a surface offering four inequivalent adsorption sites
in presence of oxygen impurities~\cite{Pfnur}.
\\

In the above-mentioned Potts models, the disorder is assumed to be
quenched and to consist in uncorrelated random couplings. However,
at some point of the evolution of the system, in particular during
its preparation, or at a much larger time scale, impurities may
diffuse in the sample and thermalize. If there exists an interaction
between them, their equilibrium configurations may display long-range
correlations. Renormalization Group studies of the $\phi^4$ model
showed that disorder correlations with an exponential or a fast
algebraic decay do not affect the critical behavior. In contrast,
slowly decaying disorder correlations, slower than $r^{-2/\nu}$ where
$\nu$ is the correlation length exponent of the pure model, bring the
system towards a new RG fixed point, distinct from the pure one and
from the one associated to uncorrelated or short-range
disorder~\cite{Weinrib}. These predictions were confirmed by Monte
Carlo simulations of the Ising model~\cite{IMCorrele,IMCorrele2}
and by RG calculations directly in dimension $d=2$~\cite{Yurko}.
\\

In the recent years, Monte Carlo simulations of the Potts model
with very slowly decaying correlations between the random couplings
led to results that do not fit in this well-accepted picture~\cite{EPL,PRE}.
Several intriguing features, already known in other models but not
in the short-range random Potts model, were observed.
First, the magnetic susceptibility displays a power-law divergence with
the lattice size, not only at the critical point, but in a finite range
of temperatures around the self-dual critical point. Such a region of
the phase diagram, known as a Griffiths phase~\cite{Griffiths}, was
first observed in the McCoy-Wu model~\cite{McCoy,McCoy2} or, equivalently
in the extreme anisotropic limit, in the random quantum Ising chain in
a transverse field~\cite{DFisher95}. It is explained as the consequence
of the existence, though with an exponentially small probability, of
macroscopically large clusters with a high concentration of strong
(resp. weak) couplings. These clusters can order earlier (later)
than the rest of the system, i.e. already in the paramagnetic (ferromagnetic)
phase~\cite{Vojta}.
Second, a violation of the hyperscaling relation $(\gamma+2\beta)/\nu=d$
was reported. Such a violation exists in pure models above their upper
critical dimension, and, for different reasons, in the classical 3D Ising
model in a random field~\cite{SchwartzSoffel}. In the latter,
the origin of the hyperscaling violation is found in the different
algebraic decay of typical and average spin-spin correlations. The
same mechanism was proposed in the long-range random Potts model.
Third, the new universality class does not seem to depend on the
number of states $q$ of the Potts model while in the case of
uncorrelated disorder a dependence of the magnetic scaling dimension
$x_\sigma=\beta/\nu$ on $q$ was unambiguously observed numerically.
The independence on the number of states $q$ is also found in the
1D random quantum Potts model~\cite{Senthil}.
Moreover, the numerical estimates of the critical exponents are
remarkably stable with the disorder strength and no sign of
cross-over could be distinguished.
\\

In this paper, the 2D classical Potts model with slowly-decaying
disorder correlations is considered. In contrast to previous studies,
the numerical calculations are performed in the limit of an
infinite disorder. After a short presentation of the model,
the magnetic scaling dimension is estimated numerically for different
disorder correlations in section~\ref{sec2}. In section~\ref{sec3},
the relevance of a large but finite disorder is investigated.
A conclusion follows.

\section{Definition of the model}
\label{sec1}
The 2D classical $q$-state Potts model is considered on the square
lattice. The Hamiltonian is~\cite{Potts}
   \be -\beta{\cal H}=\sum_{(i,j)\in E} J_{ij}\delta_{s_i,s_j},
   \hskip 1truecm s_i\in\{0,\ldots,q-1\}\label{Potts}\ee
where the sum extends over the set $E$ of pairs of neighboring sites
of the lattice. The Ising model is recovered when $q=2$.
We are interested in the case where the exchange couplings are
random and display algebraic correlations with the distance:
  \be\overline{J_{ij}J_{kl}}-\overline{J_{ij}}\ \!\overline{J_{kl}}
  \sim |\vec r_{ij}-\vec r_{kl}|^{-a}.\ee 
In previous studies, it was found convenient to generate these
coupling configurations by performing a Monte Carlo simulation of
another lattice spin model, the isotropic Ashkin-Teller model.
Its Hamiltonian~\cite{AshkinTeller}
   \be -\beta {\cal H}_{\rm AT}=\sum_{(i,j)\in E}\big[
       J_{\rm AT}(\sigma_i\sigma_j+\tau_i\tau_j)
       +K_{\rm AT}\sigma_i\sigma_j\tau_i\tau_j\big],\quad
     \sigma_i=\pm 1,\tau_i=\pm 1.\ee
is invariant under the two global ${\mathbb{Z}_2}$ transformations:
  \be \sigma_i\longrightarrow -\sigma_i, \hskip 1truecm
  (\sigma_i,\tau_i)\longrightarrow (-\sigma_i,-\tau_i)\ee
As a consequence, the phase diagram displays three phases:
a mixed, or Baxter, phase where both symmetries are spontaneously
broken, a ferromagnetic phase where the two Ising copies are
ordered but not correlated between them and a paramagnetic phase.
Magnetization and polarization
   \be M=\sum_{i\in V}\sigma_i,\hskip 1truecm
   P=\sum_{i\in V} \sigma_i\tau_i\ee
are order parameters for the phase transitions between these phases.
$V$ is the set of lattice sites. Interestingly for our purpose, the
two symmetries are simultaneously broken along a critical line which
is known exactly by self-duality arguments~\cite{Fan}:
   \be \sinh 2J_{\rm AT}=e^{-2K_{\rm AT}}\ee
While the magnetic scaling dimension $x_\sigma=1/8$ is constant
along the line, the polarization scaling dimension is given by
   \be x_{\sigma\tau}={1\over 8-4y}\ee
where the parametrization
   \be\cos{\pi y\over 2}={1\over 2}\Big[e^{4K_{\rm AT}}-1\Big]\ee
was introduced. Along the self-dual critical line,
polarization-polarization correlations decay algebraically as
   \be\overline{\sigma_i\tau_i\sigma_j\tau_j}
   \sim |\vec r_i-\vec r_j|^{-2x_{\sigma\tau}}\ee
while the average polarization density $\overline{\sigma_i\tau_i}$
vanishes.
\\

The procedure to generate correlated random couplings for the Potts
model is the following: Monte Carlo simulations of an auxiliary
Ashkin-Teller model are performed at various points $y$ of its
self-dual critical line. Statistically uncorrelated spin configurations
are sampled by throwing away a number of Monte Carlo steps several
times larger than the autocorrelation time. These spin configurations
are transformed into coupling configurations:
   \be J_{ij}={J_1+J_2\over 2}+\sigma_i\tau_i{J_1-J_2\over 2},
   \quad \forall (i,j)\in E.\label{constr}\ee
On each site, one horizontal and one vertical coupling take the same
value. This should not have any relevant effect on the physics at large
distance. For each coupling configurations, a Monte Carlo simulation of
the Potts model is performed. The construction of the random
couplings (Eq.~\ref{constr}) ensures that the disorder correlations
will decay algebraically as the polarization-polarization correlations
of the auxiliary Ashkin-Teller model. The exponent $a$ of disorder
correlation is given by $a=2x_{\sigma\tau}$. For the random Potts
model, averages are computed over both thermal and disorder
fluctuations:
   \be \overline{\langle X\rangle}={1\over{\cal Z}_{\rm AT}}
   \sum_{\{\sigma,\tau\}} \Big[{1\over{\cal Z}}\sum_{\{s\}}
     X[s]e^{\sum_{(i,j)} J_{ij}[\sigma,\tau]\delta_{s_i,s_j}}\Big]
   e^{-\beta H_{\rm AT}[\sigma,\tau]}\ee
where the brackets stands for the average over thermal fluctuations
for a given disorder configuration and the overline for the average
over disorder.

\begin{figure}
\resizebox{0.30\columnwidth}{!}{\includegraphics{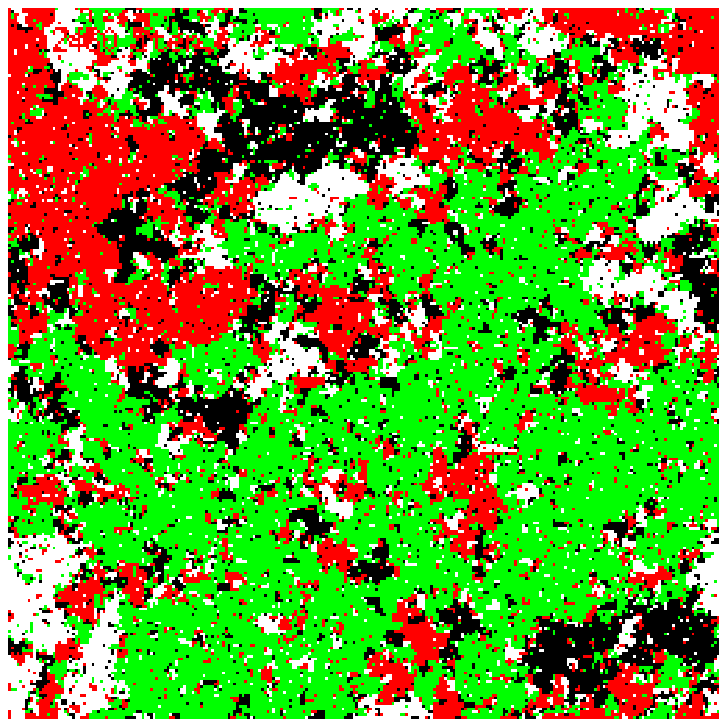}}
\resizebox{0.30\columnwidth}{!}{\includegraphics{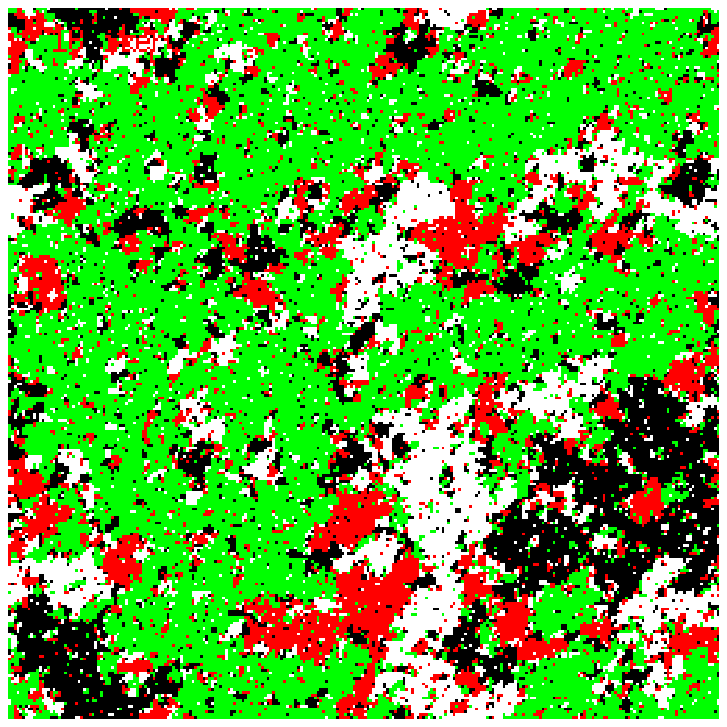}}
\resizebox{0.30\columnwidth}{!}{\includegraphics{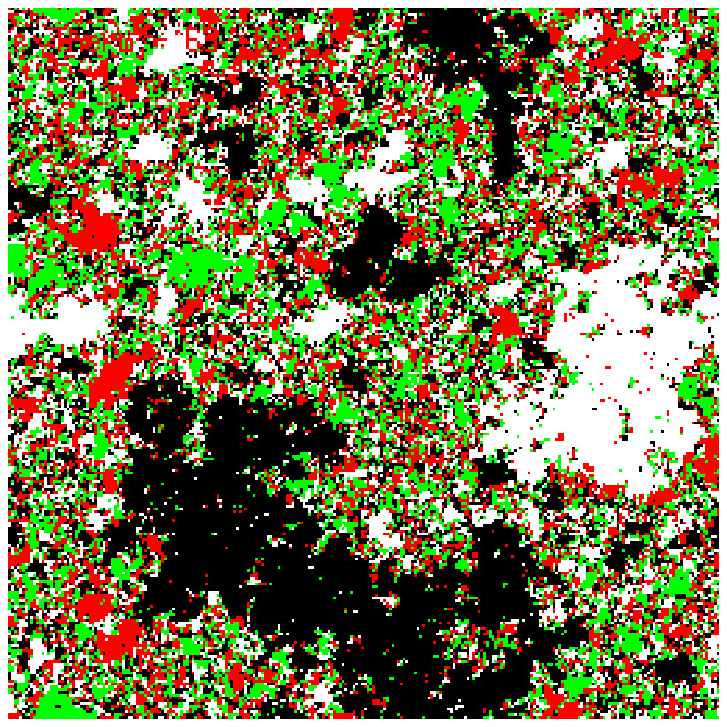}}
\caption{Typical spin configurations of the critical 4-state Potts model
  with uniform coupling (left), uncorrelated random couplings (center) and
correlated couplings (right). In the last two cases, the disorder strength
is $J_1/J_2=4$.}
\label{fig0}
\end{figure}

Typical spin configurations of the Potts model at its self dual critical
point are presented on figure~\ref{fig0}. In contrast to the case of
uncorrelated disorder, the typical spin configurations display
coexisting paramagnetic and ferromagnetic clusters when random couplings
are correlated. Paramagnetic (resp. ferromagnetic) clusters are favored
by a high density of weak (resp. strong) couplings. Thermal fluctuations
are essentially limited to the interior of each cluster and do not
induce any significant fluctuation of their boundaries. Even though the
disorder strength is relatively small ($r=J_1/J_2=4$ in the configurations
presented on the figure), the Potts model behaves as in the
limit of infinite disorder. The spin configurations of the Potts model
are highly correlated with the disorder realization and, subsequently
with the polarization configuration of the auxiliary Ashkin-Teller
model. The coexistence of paramagnetic and ferromagnetic clusters does
not imply that the random Potts model undergoes a first-order phase
transition. The Ashkin-Teller model being critical and symmetric
under polarization reversal, clusters of all sizes, both with strong
and weak couplings, are present in each disorder configuration.
Each cluster will undergo a sharp ferromagnetic-to-paramagnetic
transition at a different temperature so the average magnetization
curve will be much smoother than the curve of a single cluster. The
argument is similar to the Imry-Wortis criterion~\cite{ImryWortis}.
More important, the criticality of the Ashkin-Teller model ensures
that, in the thermodynamic limit, a percolating cluster is present
in any polarization configuration. As a consequence, a spanning
cluster of either strong or weak couplings is present in each
disorder configuration. In the former case, long-range ferromagnetic
order will be induced in the Potts model.

\section{Magnetic scaling dimension}
\label{sec2}
In this section, the magnetic scaling dimension of the long-range
random Potts model is estimated in the infinite-disorder limit.
Magnetization is more easily computed in the case of the Ising model,
i.e. $q=2$. The Hamiltonian becomes, up to a constant term,
    \be-\beta H=\sum_{(i,j)} K_{ij}s_is_j,\quad s_i\in\{+1,-1\}\ee
where $K_{ij}=J_{ij}/2$ is half of the coupling of the original
Potts model. For a given disorder realization, the partition function
reads
     \be {\cal Z}[K]=\sum_{\{s\}} e^{\sum_{(i,j)} K_{ij}s_is_j}\ee
and the average square magnetization is computed as
     \be \langle M^2\rangle={1\over{\cal Z}[K]}\sum_{\{s\}}
     \big(\sum_i s_i\big)^2e^{\sum_{(i,j)} K_{ij}s_is_j}\ee
The average over disorder leads to
   \be \overline{\langle M^2\rangle}=\int_{\R^{|E|}} {1\over{\cal Z}[K]}
   \sum_{\{s\}}\big(\sum_i s_i\big)^2e^{\sum_{(i,j)} K_{ij}s_is_j}\wp[K]
   \prod_{(i,j)} dK_{ij}\ee
As discussed in the previous section, the disorder configurations
$\{K_{ij}\}$ are either random or determined from the spin configurations
$\{\sigma,\tau\}$ of an auxiliary spin model, the Ashkin-Teller model.
In both cases, the random couplings can take only two values, $K_1$ or $K_2$.

In the limit of an infinite disorder, $K_1\rightarrow +\infty$
and $K_2\rightarrow 0$, neighboring spins connected by a coupling $K_1$
are frozen in the same state. The partition function is reduced to the
number of ways that the connected clusters of the graph formed by the
strong bonds can be decorated with Ising spins. Given a disorder
configuration $\{K_{ij}\}$, the partition function tends towards
    \be {\cal Z}[K_{ij}]=\sum_{\{s\}} e^{\sum_{(i,j)}K_{ij}s_is_j}
    \sim e^{|E_1|K_1}\sum_{\{s\}} \prod_{(i,j)\in E_1}
    \delta_{s_i,s_j}=e^{|E_1|K_1}\times 2^{C[E_1]}\ee
where $E_1=\{(i,j)/K_{ij}=K_1\}\subset E$ is the set of strong bonds,
$|E_1|=\sum_{(i,j)}\delta(K_{ij}-K_1)$ the number of strong bonds
and $C[E_1]$ the number of connected clusters in the graph $E_1$ formed
by the strong bonds. The free energy is
    \be f[K_{ij}]=-\ln{\cal Z}[K_{ij}]
    =-|E_1|K_1-C[E_1]\ln 2\ee
where $|E_1|K_1$ is an energy term while $C[E_1]\ln 2$ is the entropy
associated to the number of spin decorations of the connected
clusters of the graph $E_1$. Note also when $K_{ij}$ is determined by
$\sigma_i\tau_i$, $|E_1|$ and $C[E_1]$ are functions of the Ashkin-Teller
spin configurations $\{\sigma,\tau\}$. $|E_1|$ is related to the total
polarization $P=\sum_{i\in V} \sigma_i\tau_i$ by $|E_1|=|V|+P$ where $|V|$
is the number of lattice sites.
\smallskip

The second moment of magnetization is, for a given disorder configuration,
    \ba \langle M^2\rangle_{E_1}&=&{1\over {\cal Z}[K_{ij}]}
    e^{|E_1|K_1}\sum_{\{s\}} \big(\sum_i s_i\big)^2\prod_{(i,j)\in E_1}
    \delta_{s_i,s_j}\nonumber\\
    &\sim &2^{-C[E_1]}\sum_{\{s\}} \big(\sum_i s_i\big)^2\prod_{(i,j)\in E_1}
    \delta_{s_i,s_j}\label{eqM2a}\ea
Numerically, this quantity is evaluated in the following way:
first, the graph of strong bonds is identified and the connected clusters
are labeled. The number of sites of each clusters is determined.
Since spins are frozen in the same state in a cluster, the total
magnetization reads
   \be M[E_1]=\sum_{\alpha=1}^{C(E_1)} N_\alpha s_\alpha\ee
where $N_\alpha$ is the number of spins in the $\alpha$-th cluster and
$s_\alpha\in\{+1,-1\}$ is the value of all spins in this cluster. When
averaged over all possible spin configurations, the total magnetization
$M$ vanishes. In contrast
   \be M^2[E_1]=\Big(\sum_{\alpha=1}^{C(E_1)} N_\alpha s_\alpha\Big)^2
   =\sum_\alpha N_\alpha^2+\sum_{\alpha,\beta\ne\alpha} N_\alpha N_\beta
   s_\alpha s_\beta\ee
and, since spins are uncorrelated, i.e. $\langle s_\alpha s_\beta\rangle=0$
when $\alpha\ne\beta$, the last term vanishes and the second moment reads
   \be \langle M^2\rangle_{E_1}=\sum_\alpha N_\alpha^2.\ee
Note that the prefactor $2^{-C[E_1]}$ appearing in Eq. \ref{eqM2a} is
precisely the number of spin decorations of the clusters so that it
cancels with the sum $\sum_{\{s\}}$. Finally, the disorder
average $\overline{\langle M^2\rangle}$ is performed.
   
\begin{figure}
\psfrag{m}[Bc][Bc][1][1]{$\overline{\langle m^2\rangle}^{1/2}$}
\psfrag{L}[tc][tc][1][0]{$L$}
\resizebox{0.75\columnwidth}{!}{\includegraphics{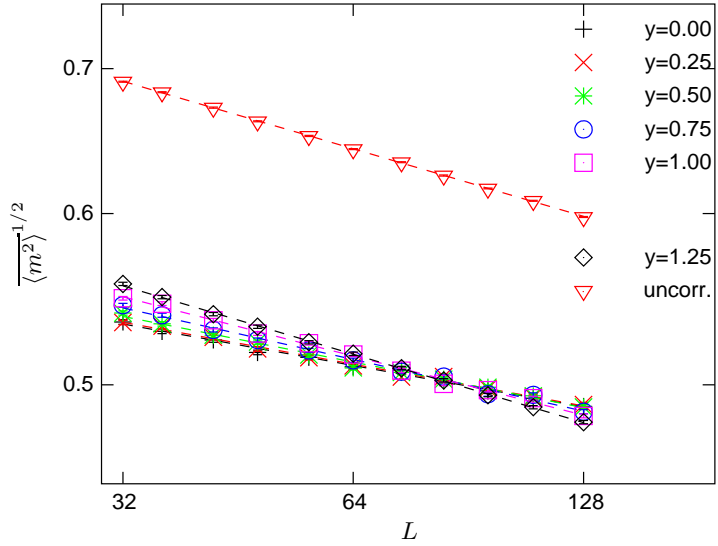} }
\caption{Finite-size scaling of the second moment of the
  magnetization density $\overline{\langle m^2\rangle}^{1/2}$
  of the 2D Ising model in the limit of infinite disorder.
  The different symbols correspond to the Monte Carlo data
  for different disorder correlations. The parameter $y$ is given
  in the legend. The case of uncorrelated disorder is also
  plotted and denoted 'uncorr.' in the legend.
  Errors bars are plotted. The dashed lines are the
  power-law fits.}
\label{fig1}
\end{figure}

On figure~\ref{fig1}, the Monte Carlo estimate of the second moment
of the magnetization density ${\overline{\langle m^2\rangle}}^{1/2}$
is plotted versus the lattice size $L$ for
different values of $y$ and for uncorrelated disorder. As discussed
above, each value of $y$ corresponds to a correlated disorder with a
different algebraic decay of disorder correlations.
The data have been averaged over 100.000 disorder configurations.
While random couplings belonging to the same disorder configuration
are correlated, there is no correlation between different disorder
configurations. Therefore, $\langle m^2\rangle$ are uncorrelated
random variables and the error on $\overline{\langle m^2\rangle}$
can be estimated from the mean square deviation.
\\

The data show a nice algebraic behavior over the full range of lattice
sizes considered. The associated critical exponents $\beta/\nu$
are reported in table~\ref{tab1}. They are in good agreement with the
values previously obtained with more standard Monte Carlo simulations
at finite disorder strength $r=K_1/K_2$~\cite{PRE}.
This supports the assumption
of a critical behavior governed by an infinite-randomness fixed point.
In contrast, the estimate of $\beta/\nu$ for uncorrelated disorder
($0.1037(4)$) is incompatible with the expected exponent for the 2D random
Ising model. Since randomness is marginally irrelevant in this case,
the exponent $\beta/\nu=1/8$ of the pure fixed point was expected.
However, we note that the estimate obtained in the limit of an infinite
disorder is close, though at the boundary of error bars, to the exponent
${\beta\over \nu}={5\over 36}\times {3\over 4}\simeq 0.10417\ldots$
of percolation. This can be understood in the following way: in the limit
$K_1\rightarrow +\infty$ and $K_2\rightarrow 0$, the random-bond Ising model
is equivalent to a diluted Ising model at zero temperature. Long-range
ferromagnetic order can exist only if the bonds percolate. The singularity
of magnetization is due to the percolation transition. In the case of
the random-bond Ising or Potts model, the percolation fixed point is
known to be unstable. It manifests itself only by cross-over
effects at small lattice sizes and strong disorder. In the simulations
presented above, the calculation is performed exactly at the percolation
fixed point. As a consequence, the exponent $\beta/\nu$ of the percolation
fixed point is measured and no cross-over effect is expected.
\\

To summarize, in the case of correlated disorder, the magnetic critical
behavior that was previously reported at finite disorder strength is the
same as in the limit of infinite strength. It is therefore governed by the
correlated percolation fixed point~\cite{Coniglio}. As far as we are aware,
only the value corresponding to the Ising model, i.e. $y=0$, has been
reported to be $0.0527(4)$ assuming $\nu=1$~\cite{Janke}. Even though no
cross-over was observed at finite disorder strength, the stability of the
fixed point remains to be shown. In the case of uncorrelated disorder, the
percolation fixed point is unstable so that, at finite disorder, and
therefore at finite temperature at the critical point, thermal fluctuations
bring the system to the random fixed point if disorder is relevant, or to
the pure fixed point otherwise.

\begin{table}
  \caption{Numerical estimates of the magnetic scaling dimension $\beta/\nu$
    for correlated disorder ($y\in [0;1.25]$) and for uncorrelated disorder
    ('uncorr.') as discussed in section~\ref{sec2}. The exponent $a$ of
    the algebraic decay of disorder correlation is given in the second column.
    The last two columns correspond to the Monte
    Carlo estimates of the exponent $\omega$ of the algebraic decay of the
    first-order corrections $C_1$ and $C_2$ of the square magnetization
    (see section~\ref{sec3} for details).}
\label{tab1}
\begin{tabular}{lllll}
\hline\noalign{\smallskip}
$y$ & $a$ & $\beta/\nu$ & Correction \\
\hline\noalign{\smallskip}
$0.00$ & $0.25$ & $0.064(2)$ & $0.135(3)$ & $0.125(3)$ \\
$0.25$ & $0.286$ & $0.064(2)$ & $0.144(3)$ & $0.134(3)$\\
$0.50$ & $0.333$ & $0.070(2)$ & $0.170(3)$ & $0.152(3)$ \\
$0.75$ & $0.4$ & $0.079(2)$ & $0.190(3)$ & $0.167(3)$ \\
$1.00$ & $0.5$ & $0.091(2)$ & $0.194(3)$ & $0.166(3)$ \\
$1.25$ & $0.667$ & $0.1050(8)$ & $0.192(3)$ & $0.161(3)$ \\
uncorr. & $2$ & $0.1037(4)$ & $0.786(2)$ & $0.788(3)$ \\
\hline\noalign{\smallskip}
\end{tabular}
\end{table}

\section{Stability of the correlated percolation fixed point}
\label{sec3}

In the previous section, the calculations were performed exactly in
the limit of an infinite disorder, i.e. at the (correlated) percolation
point. In this section, the case of a large but finite disorder is
considered. The partition function and the average square magnetization
are expanded to first order around the (correlated) percolation
point and the exponent of the first correction is estimated.
\\

Even though the Ising case $q=2$ will be considered, we start with the
Potts Hamiltonian. For a given disorder configuration $\{J_{ij}\}$, the
partition function reads
    \ba {\cal Z}[J_{ij}]=\sum_{\{s\}} e^{\sum_{(i,j)} J_{ij}\delta_{s_i,s_j}}
    =\sum_{\{s\}} \prod_{(i,j)} \big(u_{ij}\delta_{s_i,s_j}+1\big)\ea
where
    \be u_{ij}=e^{J_{ij}}-1.\ee
Introducing $E_1$ (resp. $E_2$) the set of bonds for which $J_{ij}=J_1$
(resp. $J_{ij}=J_2$), the partition function of the Potts model for a given
coupling realization reads
    \be {\cal Z}[J_{ij}]=\sum_{\{s\}}\prod_{(i,j)\in E_1}
    \big(u_1\delta_{s_i,s_j}+1\big)
    \prod_{(i,j)\in E_2}\big(u_2\delta_{s_i,s_j}+1\big)\ee
On the square lattice, each coupling configuration $J_{ij}$ can be mapped
onto a dual configuration $J_{ij}^*$ where
    \be u_{ij}u_{ij}^*=q\ee
with $u_{ij}^*=e^{J_{ij}^*}-1$. The duality transformation $J_{ij}\rightarrow
J_{ij}^*$ exchanges strong and weak couplings and therefore maps the
ferromagnetic phase onto the paramagnetic one. If a coupling configuration
$J$ and its dual $J^*$ have the same probability, i.e. $\wp[J]=\wp[J^*]$,
the model is self-dual and critical. In our case, this condition is
satisfied when $J_1=J_2^*$ because the exchange of the couplings $J_1$ and
$J_2$ is equivalent to a reversal of the polarization of the auxiliary
Ashkin-Teller model. Along the critical line, the variable $u_2=q/u_1$
can be removed:
   \be {\cal Z}[J_{ij}]=u_1^{|E_1|}\sum_{\{s\}}\prod_{(i,j)\in E_1}
   \Big(\delta_{s_i,s_j}+{1\over u_1}\big)\prod_{(i,j)\in E_2}
   \Big({q\over u_1}\delta_{s_i,s_j}+1\Big)\ee
where $|E_1|$ is the number of couplings $J_1$. When $u_1\gg u_2=q/u_1$, an
expansion in powers of $1/u_1$ is easily obtained. To first order,
the partition function is
   \be {\cal Z}[J_{ij}]\simeq u_1^{|E_1|}\sum_{\{s\}}
   \Big[\prod_{(i,j)\in E_1}\delta_{s_i,s_j}+{1\over u_1}
   \Big(\sum_{(k,l)\in E_1}\prod_{(i,j)\atop\in E_1\setminus(k,l)}\delta_{s_i,s_j}
   +q\sum_{(k,l)\in E_2}\prod_{(i,j)\atop\in E_1\cup(k,l)}\delta_{s_i,s_j}
   \Big)\Big]\ee
Each term is proportional to the number of ways a graph can be decorated
by Potts spins. Denoting $C(G)$ the number of connected clusters of a graph
$G\subset E$, the partition function reads to first order
    \ba{\cal Z}&&[J_{ij}]\simeq u_1^{|E_1|} \Big[q^{C(E_1)}+{1\over u_1}
    \Big(\sum_{(k,l)\in E_1}q^{C(E_1\setminus(k,l))}+\sum_{(k,l)\not\in E_1}
    q^{C(E_1\cup(k,l))+1}\Big)\Big]\\
    &&=u_1^{|E_1|}q^{C(E_1)}\Big[1+{1\over u_1}\Big(\sum_{(k,l)\in E_1}
      q^{C(E_1\setminus(k,l))-C(E_1)}+\sum_{(k,l)\not\in E_1}q^{C(E_1\cup(k,l))+1-C(E_1)}
      \Big)\Big]\nonumber
    \ea
The quantity $C(E_1\setminus(k,l))-C(E_1)$ vanishes for all bonds $(k,l)$
of $E_1$ that do not disconnect a cluster into two unconnected parts when
they are removed. It takes the value $+1$ otherwise, i.e. for the
so-called red bonds, or bridges, of the graph $E_1$. The quantity
$C(E_1\cup(k,l))-C(E_1)$ vanishes if adding a bond $(k,l)$ does not change
the number of clusters, i.e. if the vertices $k$ and $l$ are connected by
some path on the graph $E_1$. It takes the value $-1$ otherwise, i.e. when
the vertices $k$ and $l$ belong to different clusters of the graph $E_1$.
Numerically, the first quantity is determined by identifying the red bonds
with the Tarjan algorithm~\cite{Tarjan}. The second quantity is computed
using a labelling of the clusters during their identification.
\\
    
The same expansion is now written for the average square magnetization
of the Ising model, i.e. the case $q=2$:
   \be \langle M^2\rangle={1\over{\cal Z}[J_{ij}]}
   u_1^{|E_1|}\sum_{\{s\}}\big(\sum_i s_i\big)^2\prod_{(i,j)\in E_1}
   \Big(\delta_{s_i,s_j}+{1\over u_1}\big)\prod_{(i,j)\in E_2}
   \Big({q\over u_1}\delta_{s_i,s_j}+1\Big)\ee
To first order and performing the average over the spin decorations
of the clusters, the square magnetization reads
   \ba\langle M^2\rangle&\simeq&\langle M^2\rangle_{E_1}
   \nonumber\\
   &+&{q^{-C(E_1)}\over u_1}\Big(\sum_{(k,l)\in E_1}
   q^{C(E_1\setminus(k,l))}\langle M^2\rangle_{E_1\setminus(k,l)}
   +q\!\!\!\sum_{(k,l)\in E_2}q^{C(E_1\cup(k,l))}\langle M^2\rangle_{E_1\cup(k,l)}\Big)
   \nonumber\\
   &-&{1\over u_1}\langle M^2\rangle_{E_1}
   \Big(\sum_{(k,l)\in E_1}\!\!\!q^{C(E_1\setminus(k,l))-C(E_1)}
   +\!\!\!\sum_{(k,l)\not\in E_1}\!\!\! q^{C(E_1\cup(k,l))+1-C(E_1)}\Big)
   \ea
where $\langle M^2\rangle_{E_1}$ denotes a calculation at zeroth-order
in $1/u_1$ for the graph $E_1$ along the lines presented after
Eq. \ref{eqM2a}. Finally, we get
   \ba\langle M^2\rangle\simeq\langle M^2\rangle_{E_1}
   &+&{1\over u_1}\sum_{(k,l)\in E_1}\!\!\!q^{C(E_1\setminus(k,l))-C(E_1)}
   \big(\langle M^2\rangle_{E_1\setminus(k,l)}-\langle M^2\rangle_{E_1}\big)
   \nonumber\\
   &+&{1\over u_1}\sum_{(k,l)\in E_2}q^{C(E_1\cup(k,l))+1-C(E_1)}
   \big(\langle M^2\rangle_{E_1\cup(k,l)}-\langle M^2\rangle_{E_1}\big)
   \label{eqM2b}
   \ea
The only strong couplings $J_{kl}$, i.e. $(k,l)\in E_1$, contributing to the
first sum are the red bonds of the graph $E_1$. Their removal provokes the
splitting of a cluster of strong couplings into two unconnected
clusters so, as discussed before, $C(E_1\setminus(k,l))-C(E_1)=1$. The
change in the total square magnetization, $\langle M^2\rangle_{E_1\setminus(k,l)}
-\langle M^2\rangle_{E_1}$, is computed numerically simply by first removing
the red bond and then relabelling the cluster starting at one edge of the
bond. The only weak couplings $J_{kl}$, i.e. $(k,l)\not\in E_1$, contributing
to the second sum of Eq. \ref{eqM2b} are those for which a different cluster
is found at the two edges. Adding this bond leads to the merging of the
two clusters so that $C(E_1\cup(k,l))+1-C(E_1)=0$. The calculation of the
change in the total square magnetization, $\langle M^2\rangle_{E_1\cup(k,l)}
-\langle M^2\rangle_{E_1}$, only requires a minimal computational effort
if the sizes of the different clusters were stored. After performing
the average over disorder configurations, the average
square magnetization can be put in the form
   \be \overline{\langle M^2\rangle}_{u_1}
   =\overline{\langle M^2\rangle}_{\infty}
   \Big[1+{1\over u_1}\big(qC_1+C_2\big)\Big]\ee
where the two correction terms are
   \be C_1={\overline{\sum_{(k,l)\in E_1}\big[
   \langle M^2\rangle_{E_1\setminus(k,l)}-\langle M^2\rangle_{E_1}\big]}_{\infty}
     \over \overline{\langle M^2\rangle}_{\infty}}\ee
and
   \be C_2={\overline{\sum_{(k,l)\not\in E_1}\big[
   \langle M^2\rangle_{E_1\cup(k,l)}-\langle M^2\rangle_{E_1}\big]}_{\infty}
     \over \overline{\langle M^2\rangle}_{\infty}}\ee
   
\begin{figure}
  \psfrag{Terme Corr. 1}[Bc][Bc][1][1]{$C_1$}
  \psfrag{Terme Corr. 2}[Bc][Bc][1][1]{$C_2$}
  \psfrag{L}[tc][tc][1][0]{$L$}
  \resizebox{0.5\columnwidth}{!}{\includegraphics{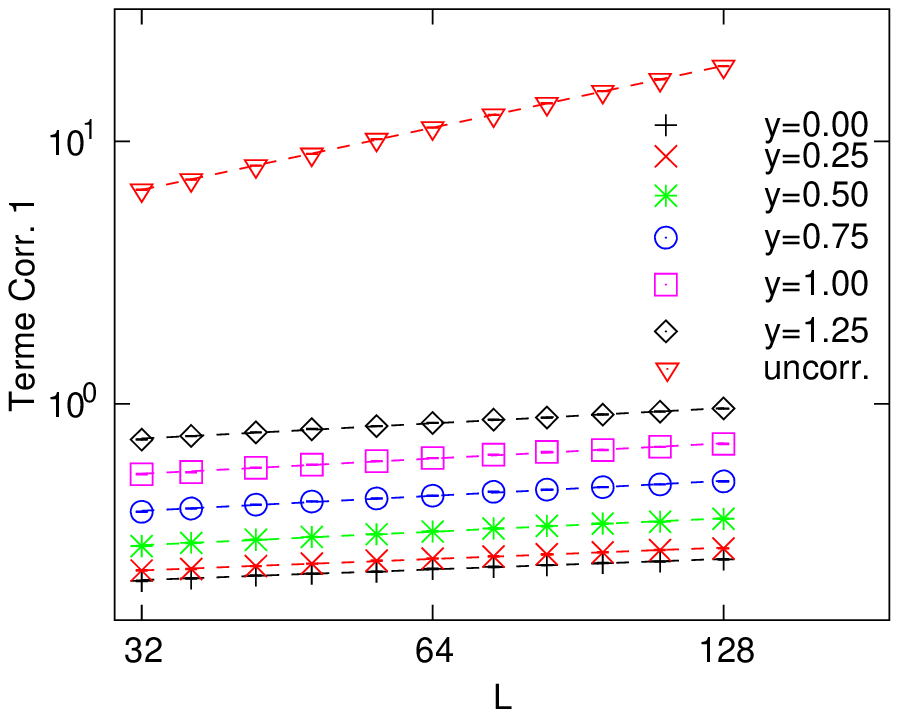}}
  \resizebox{0.5\columnwidth}{!}{\includegraphics{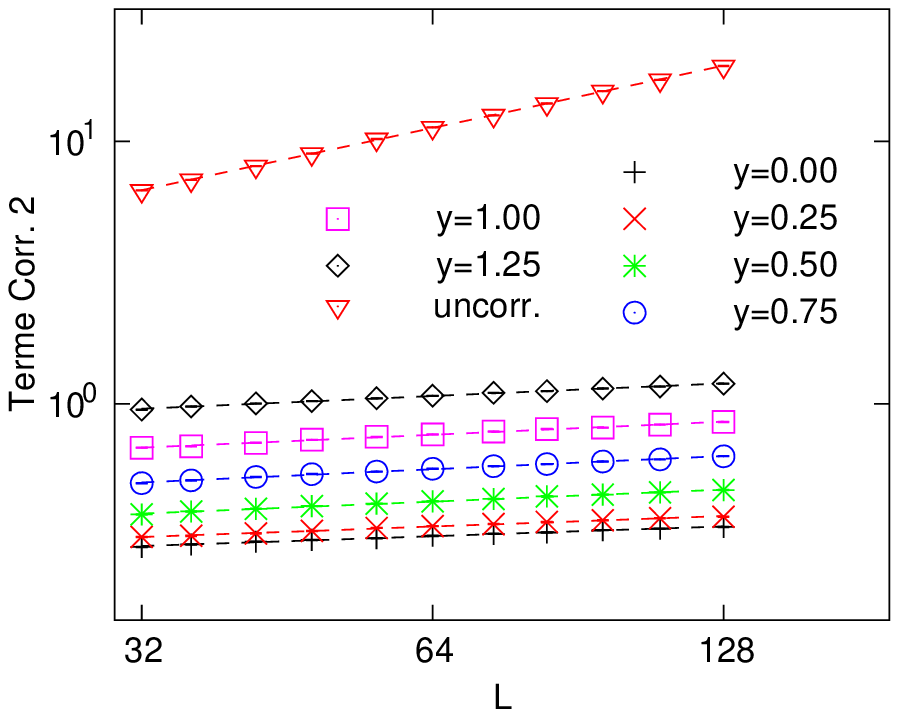}}
  \caption{Finite-size scaling of the two contributions $C_1$ (left)
    and $C_2$ (right) to the first-order
    correction to the square magnetization of the 2D Ising model in the
    limit of infinite disorder. The different symbols correspond to the
    Monte Carlo data for different disorder correlations. The parameter
    $y$ is given in the legend. The case of uncorrelated disorder is also
    plotted and denoted 'uncorr.' in the legend. Errors bars are plotted.
    The dashed lines are the power-law fits.}
\label{fig2}
\end{figure}

The two correction terms $C_1$ and $C_2$ were computed for 100.000 graphs
$E_1$ corresponding to disorder configurations $J_{ij}$, and therefore to
polarization configurations of the auxiliary Ashkin-Teller model. The data
are plotted on figure~\ref{fig2} versus the lattice size $L$. Nice power
laws are observed. Slightly larger exponents are obtained for $C_1$ than for
$C_2$ (see the estimates in table~\ref{tab1}). In the case of uncorrelated
disorder, this exponent is $\omega\simeq 0.79$. In contrast, for correlated
disorder, the exponent is much smaller but positive. There are two important
consequences: first, the correction terms grow with the lattice size which
implies that they will eventually become dominant. The correlated percolation
fixed point is therefore unstable. Second, the cross-over length $L^*$ at
which the correction becomes of order ${\cal O}(1)$ for a given $u_1$ is
much smaller in the uncorrelated case. For $L=32$ for instance, $C_1\simeq
6.55$ for uncorrelated disorder while $C_1\simeq 0.73$ for $y=1.25$. Assuming
a correction exponent $\omega\simeq 0.2$, $C_1$ will be equal to $6.55$
only at a lattice size $L\simeq 2.3.10^6$! The situation is even worse in the
case $y=0$ for which a lattice size $L\simeq 2.5.10^{12}$ is needed for
$C_1$ to reach the value $6.55$. Such huge lattice sizes cannot be reached
by Monte Carlo simulations. This explains why the cross-over was not
observed in previous Monte Carlo simulations at finite disorder strength.

\section{Conclusion}
By performing calculations explicitly in the limit of an infinite disorder
strength, it was shown that the magnetic scaling dimensions previously reported
for the random Potts model with slowly-decaying disorder correlations
correspond to the correlated percolation fixed point. The analysis
of the first-order corrections at finite disorder strength revealed that
this fixed point is unstable. However, the cross-over length associated to
these corrections is found to be much larger, by several orders of magnitude,
than in the uncorrelated case. Therefore, the true critical behavior cannot
be reached by Monte Carlo simulations.
Nevertheless, it is interesting that this model reproduces features
of an infinite-disorder fixed point even though the latter is not stable
in this case. One may hope that it will help in a better understanding of
the stability of such fixed points.

\section{Acknowledgments}
The author would like to congratulate Wolfhard Janke at the occasion of
his 60th birthday and gratefully thanks the organizers of the workshop
dedicated to this event.

\end{document}